\documentclass[11pt,a4paper]{article}
\usepackage{amsmath}
\usepackage[mathcal]{eucal}
\usepackage{latexsym}
\usepackage{amssymb}
\begin{titlepage}
\title{A brief survey of the renormalizability of four dimensional gravity for generalized Kodama states.}
\author{Eyo Eyo Ita III}
\medskip
\input amssym.def
\input amssym.tex

\begin{document}
\maketitle
\bigskip
\centerline{Department of Applied Mathematics and Theoretical Physics} 
\smallskip
\centerline{Centre for Mathematical Sciences, University of Cambridge, Wilberforce Road}
\smallskip
\centerline{Cambridge CB3 0WA, United Kingdom}
\smallskip
\centerline{eei20@cam.ac.uk} 

\bigskip

\begin{abstract}
We continue the line of research from previous works in assessing the suitability 
of the pure Kodama state both as a ground state for the generalized Kodama states, as well as characteristic of a good semiclassical limit of general relativity.  We briefly introduce the quantum theory of fluctuations about DeSitter spacetime, which enables one to examine some perturbative aspects of the state.  Additionally, we also motivate the concept of the cubic tree network, which enables one to view the generalized Kodama states in compact form as a nonlinear transformation of the pure Kodama states parametrized by the matter content of the proper classical limit.  It is hoped that this work constitutes a first step in addressing the nonperturbative renormalizability of general relativity in Ashtekar variables.  Remaining issues to address, including the analysis of specific matter models, include finiteness and normalizability of the generalized Kodama state as well as reality conditions on the Ashtekar variables, which we relegate to separate works.
\end{abstract}
\end{titlepage}

\section{Introduction: Stability of the pure Kodama state}

\noindent
This work continues the line of research from \cite{EYO}, \cite{EYOFULL}.  The main question that would like to formulate is (i) whether or not the pure Kodama state $\Psi_{Kod}$ is a good ground state of general relativity about which quantum fluctuations constitue a renormalizable theory.  The second main question we would like to analyse is concerning the relation between the pure and the generalized Kodama states with respect to the vacuum state of quantum general relativity: (ii) Is $\Psi_{GKod}$ in any sense an excited version of $\Psi_{Kod}$, or is it an independent ground state of the gravity-matter system?  As a corollary to (ii), does each model for which a $\Psi_{GKod}$ can be constructed consitute an additional vacuum state of general relativity?  In \cite{SOO}, Chopin Soo and Lee Smolin present the hypothesis for the treatment of matter fields as a perturbation about DeSitter space satisfying a Schr\"odinger equation for small fluctuations.  In \cite{LINKO}, Smolin and Friedel expand $\Psi_{Kod}$ in gravitions about an abelian theory.  However, since the generalized Kodama states are designed to incorporate the matter effects to all orders and to enforce the proper semiclassical limit, we expect it to be the case that $\Psi_{GKod}$ is to all orders in the expansion nonperturbatively related to $\Psi_{Kod}$.  In this case, one should be able to find a discrete transformation between that maps $\Psi_{Kod}$ into $\Psi_{GKod}$ and more generally, a discrete transformation amongst the generalized Kodama states for different models.\par
\indent
The manner in which we address this transformation is to view $\Psi_{Kod}$ as being invariant under a symmetry which is broken due to the presence of matter fields.  When one views the effect of the matter fields in terms of backreactions on DeSitter spacetime, then one can see more clearly the link from $\Psi_{GKod}$ to the semiclassical limit below the Planck scale.  We provide in this work a brief synopsis of the required transformation in terms of tree networks, and then briefly comment in the discussion on the implications for nonperturbative renormalizability in the Ashtekar variables.\par
\indent
The layout of this paper is as follows.  In section 2 we review the developments which cast the pure Kodama state into a perspective suitable for posing the question of a stable ground state.  In section 3 we discuss in detail the effects and the interpretation of incorporating matter fields, in the good semiclassical limit below the Planck scale, into the fully extrapolated theory of quantized gravity.  In section 4 we briefly introduce the quantum theory of fluctuations on DeSitter spacetime and the general structures required.  In section 5 we introduce the concept of the tree network, which can be seen as the application of Feynman diagrammatic techniques to the solution of the constraints.  We then show how the networks implement the discrete transformation amongst generalized Kodama states.
We argue for the interpretation of general relativity as a renormalizable theory due to its tree network structure when expressed in Ashtekar variables.

\section{The Pure Kodama state as a ground state of quantized gravity}

\indent
To recapitulate the results of \cite{EYO},\cite{EYOFULL}, when one quantizes Einstein's general relativity in Ashtekar variables subject to the CDJ Ansatz and the semiclassical-quantum correspondence, one obtains a set of nine conditions on the nine elements of the CDJ matrix $\Psi_{ae}$, of the general form\footnote{The mixed partials condition, a link from quantized gravity to the semiclassical limit below the Planck scale, should be incorporated as well into the solution of the system as a consistency condition stemming from the canonical quantization relations. From the dimensionally expanded view, one then solves a 9+N by 9+N system of equations.  Also, the constraints of relativity bear a distinct relation in structure to the classical equations of motion for 
scalar $\phi^4$ theory $(\square+m^2)\phi+{g \over {2!}}\phi^2+{\lambda \over {4!}}\phi^3=0$, as well as to Yang--Mills theory, two renormalizable theories. We argue later for the implication of general relativity as a renormalizable theory in Ashtekar variables based upon this structure.}

\begin{eqnarray}
\label{SYSKOD2}
C_{ab}=I_{ab}^{cd}\Psi_{cd}-GQ_{ab}+\Lambda{I}_{ab}^{cdef}\Psi_{cd}\Psi_{ef}+\Lambda^2{I}_{ab}^{cdefgh}\Psi_{cd}\Psi_{ef}\Psi_{gh}=0,
\end{eqnarray}

\noindent
where $\Lambda$ is the cosmological constant.  To asses the viability of the pure Kodama 
state $\Psi_{Kod}$ as a ground state of the theory, one may expand the CDJ matrix in terms of deviations from DeSitter spacetime

\begin{eqnarray}
\label{CDDJ}
\Psi_{ae}=-\Bigl({6 \over \Lambda}\delta_{ae}+\epsilon_{ae}\Bigr)
\end{eqnarray}

\noindent
and then substitute (\ref{CDDJ}) into (\ref{SYSKOD2}).  The constraints can be written in the following form in terms of the CDJ deviation 
matrix $\epsilon_{ae}$ which encodes the departure from DeSitter spacetime.  In the absence of matter fields the quantum constraints of general relativity in Ashtekar variables correspond to the set of nine equations $C^{gh}$ in nine unknowns $\epsilon_{ae}$

\begin{eqnarray}
\label{FORM}
C^{gh}[\epsilon_{ae}]=O^{ghae}\epsilon_{ae}+\Lambda{I}^{ghaebf}\epsilon_{ae}\epsilon_{bf}
+\Lambda^2{E}^{ghabcdef}\epsilon_{ad}\epsilon_{be}\epsilon_{cf}=0.
\end{eqnarray}

\noindent
where the operator $O^{ghae}$ is in general a nine by nine matrix of integro-differential operators, which can be written in the compact form

\begin{eqnarray}
\label{FORM1}
O^{ghae}=\sum_{d=1}^{3}\epsilon_{[aed]}\epsilon_{dgh}
+\sum_{d=1}^{3}G^{gh}_{d}\epsilon_{(aed)}+\eta^{ae}\delta_{g1}\delta_{h1}+\partial^{ae}\delta_{g2}\delta_{h2}+\Delta^{ae}\delta_{g3}\delta_{h3}
\end{eqnarray}

\noindent
where the notation $\epsilon_{(aed)}$ signifies the even permutations of the epsilon tensor $(\epsilon_{123},\epsilon_{231},\epsilon_{312})$, used to implement the Gauss' law constraint and $\epsilon_{[aed]}$ signifies the odd permutations, $(\epsilon_{213},\epsilon_{321},\epsilon_{132})$ are used to implement the diffeomorphism constraint.  The meaning of the operators in (\ref{FORM1}) is as follows.  The operator $\epsilon_{aed}$ implements the diffeomorphism constraint while $G^{ae}_d$ implements the Gauss' law constraint.  $\partial^{ae}$ is the functional divergence and $\Delta^{ae}$ is the functional Laplacian, while $\eta^{ae}$ implements the trace of the CDJ deviation matrix.  The matrix $O^{ab}_{cd}$, fully written out, is given by \cite{EYOFULL}

\begin{displaymath}
O_{ab}^{cd}=
\left(\begin{array}{ccccccccc}
-1 & 0 & 0 & 1 & 0 & 0 & 0 & 0 & 0\\
0 & -1 & 0 & 0 & 1 & 0 & 0 & 0 & 0\\
0 & 0 & -1 & 0 & 0 & 1 & 0 & 0 & 0\\
(\hat{G}_1)^{21} & (\hat{G}_1)^{32} & (\hat{G}_1)^{13} & (\hat{G}_1)^{12} & (\hat{G}_1)^{23} & (\hat{G}_1)^{31} & (\hat{G}_1)^{11} & (\hat{G}_1)^{22} & (\hat{G}_1)^{33}\\ 
(\hat{G}_2)^{21} & (\hat{G}_2)^{32} & (\hat{G}_2)^{13} & (\hat{G}_2)^{12} & (\hat{G}_2)^{23} & (\hat{G}_2)^{31} & (\hat{G}_2)^{11} & (\hat{G}_2)^{22} & (\hat{G}_2)^{33}\\ 
(\hat{G}_3)^{21} & (\hat{G}_3)^{32} & (\hat{G}_3)^{13} & (\hat{G}_3)^{12} & (\hat{G}_3)^{23} & (\hat{G}_3)^{31} & (\hat{G}_3)^{11} & (\hat{G}_3)^{22} & (\hat{G}_3)^{33}\\ 
0 & 0 & 0 & 0 & 0 & 0 & 1 & 1 & 1\\
\hat{\partial}^{21} & \hat{\partial}^{32} & \hat{\partial}^{13} & \hat{\partial}^{12} & \hat{\partial}^{23} & \hat{\partial}^{31} 
& \hat{\partial}^{11} & \hat{\partial}^{22} & \hat{\partial}^{33}\\ 
\hat{\Delta}^{21} & \hat{\Delta}^{32} & \hat{\Delta}^{13} & \hat{\Delta}^{12} & \hat{\Delta}^{23} & \hat{\Delta}^{31} 
& \hat{\Delta}^{11} & \hat{\Delta}^{22} & \hat{\Delta}^{33}\\ 
\end{array}\right).
\end{displaymath}

\noindent
Also, for the pure Kodama state the following definitions apply

\begin{eqnarray}
\label{FORM2}
I^{ghabef}={1 \over 6}\delta_{g1}\delta_{h1}\bigl(\delta_{ae}\delta_{bf}-\delta_{af}\delta_{be}\bigr)
+{1 \over 8}\delta_{g2}\delta_{h2}\partial^{eb}_{af};\nonumber\\
E^{ghabcdef}={1 \over {72}}\delta_{g3}\delta_{h3}\epsilon_{abc}\epsilon_{def}.
\end{eqnarray}

\noindent
The unique solution to the system (\ref{FORM}), its critical point, is given by $\epsilon_{ae}=0$ which corresponds to a self-dual spacetime.  This is DeSitter spacetime and which is manifested at the semiclassical level via the pure Kodama state $\Psi_{Kod}$.  The significance of the matrix $O_{ab}^{cd}$ is that it acts as a kind of kinetic term in 
(\ref{FORM}) for the $\epsilon_{ae}$ field.  Hence its inverse should serve as a kind of functional propagator for the 
theory.  Equation (\ref{FORM}) can be rearranged into the following recursion relation

\begin{eqnarray}
\label{FORMM}
\epsilon_{mn}=-\Lambda(O^{-1})_{mngh}{I}^{ghaebf}\epsilon_{ae}\epsilon_{bf}-\Lambda^2(O^{-1})_{mngh}{E}^{ghabcdef}\epsilon_{ad}\epsilon_{be}\epsilon_{cf}.
\end{eqnarray}

\noindent
Equation (\ref{FORM}) can be iterated by substituting all occurences of the CDJ deviation matrix for itself on the right hand side.  This leads to an infinite series expansion in powers of $\Lambda$.  The first few terms of the series read

\begin{eqnarray}
\label{FORMM1}
\epsilon_{mn}=
\hbox{lim}_{N_k\rightarrow\infty}
\biggl[\Lambda^{N_1}
\Bigl((O^{-1})_{mngh}{I}^{ghaebf}(O^{-1})_{aea^{\prime}e^{\prime}}(O^{-1})_{bfb^{\prime}f^{\prime}}
I^{a^{\prime}e^{\prime}g^{\prime}h^{\prime}}I^{b^{\prime}f^{\prime}m^{\prime}n^{\prime}}...\Bigr)\nonumber\\
+\Lambda^{N_2}\Bigl((O^{-1})_{mngh}{I}^{ghaebf}(O^{-1})_{aea^{\prime}e^{\prime}}(O^{-1})_{bfb^{\prime}f^{\prime}}
{E}^{a^{\prime}e^{\prime}a_1b_1c_1d_1e_1f_1}{E}^{b^{\prime}f^{\prime}a_2b_2c_2d_2e_2f_2}...\Bigr)\nonumber\\
+\Lambda^{N_3}\Bigl(
(O^{-1})_{mngh}{E}^{ghabcdef}(O^{-1})_{ada^{\prime}d^{\prime}}(O^{-1})_{beb^{\prime}e^{\prime}}(O^{-1})_{cfc^{\prime}df^{\prime}}...\Bigr)\biggr]
\end{eqnarray}

\noindent
Though (\ref{FORMM1}) contains an infinite number of terms, the terms fall into two main categories.  Think of the matrix $(O^{-1})_{mngh}$ as a propagator for the linearized theory of $\Psi_{Kod}$ and represent it as a straight line on a graph.  There are only two types of vertices, defined in (\ref{FORM2}).  There is 
the ${I}^{ghaebf}$ vertex, which is associated with a factor of $\Lambda$, and the ${E}^{ghabcdef}$ vertex, associated with a factor of $\Lambda^2$, as indicated 
in (\ref{FORM}).  One can view the expansion (\ref{FORMM1}) conveniently in terms of a network in which the lines emanate from and terminate on these two different types of vertices in all possible ways.\par
\indent  
Note that since the constraint equations (\ref{FORMM}) in the absence of matter fields do not contain a source term, the network continues to infinity without any breaks, accumulating powers of $\Lambda$ from each vertex.  Since each term of the diagram is infinitely connected, there is a corresponding infinity of powers 
of $\Lambda$.  So the solution is given by $\epsilon_{mn}\propto\Lambda^{\infty}$.  This places an upper bound on the cosmological constant, as for any solution corresponding to $\epsilon_{mn}\neq{0}$ one would need $\Lambda<1$ as a necessary condition for convergence of the sequence.\footnote{We will see that the existence of matter fields coupled to gravity will place tighter bounds on the value of the cosmological constant.}  But then $\epsilon_{mn}$ would converge to zero, since any deviation from the CDJ matrix $\Psi_{ae}=-6\Lambda^{-1}\delta_{ae}$ will be exterminated by the infinite powers of $\Lambda$.  The solution $\epsilon_{mn}=0$ is a self consistent solution to the constraints, and the unique nonperturbatively exact solution corresponding to the pure Kodama state $\Psi_{Kod}$.\par
\indent
In order to solve the constraints, one must be able to find the propagator for the theory.  This amounts to computing $(O^{-1})_{mngh}$, the inverse of the kinetic operator.  Given that the matrix elements of the kinetic operator are in general noncommuting operators, one must exercise extreme care when attempting to invert it with regard for operator ordering.  One possible method of inversion is to partition the matrix into three by three blocks 

\begin{displaymath}
O_{ab}^{cd}=
\left(\begin{array}{ccc}
\boldsymbol{D}_1 & \boldsymbol{D}_2 & \boldsymbol{D}_3\\
\boldsymbol{G}_1 & \boldsymbol{G}_2 & \boldsymbol{G}_3\\
\boldsymbol{H}_1 & \boldsymbol{H}_2 & \boldsymbol{H}_3\\
\end{array}\right).
\end{displaymath}

\noindent 
where the blocks denote their corresponding constraints.

\begin{displaymath}
\boldsymbol{D}_1=
\left(\begin{array}{ccc}
-1 & 0 & 0 \\
0 & -1 & 0 \\
0 & 0 & -1 \\
\end{array}\right);~~
\boldsymbol{D}_2=
\left(\begin{array}{ccc}
1 & 0 & 0 \\
0 & 1 & 0 \\
0 & 0 & 1 \\
\end{array}\right);~~
\boldsymbol{D}_3=
\left(\begin{array}{ccc}
0 & 0 & 0 \\
0 & 0 & 0 \\
0 & 0 & 0 \\
\end{array}\right)
\end{displaymath}

\noindent
Correspond to the diffeomorphism constraints in the absence of matter fields.

\begin{displaymath}
\boldsymbol{G}_1=
\left(\begin{array}{ccc}
(\hat{G}_1)^{21} & (\hat{G}_1)^{32} & (\hat{G}_1)^{13} \\
(\hat{G}_2)^{21} & (\hat{G}_2)^{32} & (\hat{G}_2)^{13} \\
(\hat{G}_3)^{21} & (\hat{G}_3)^{32} & (\hat{G}_3)^{13} \\
\end{array}\right);~~
\boldsymbol{G}_2=
\left(\begin{array}{ccc}
(\hat{G}_1)^{12} & (\hat{G}_1)^{23} & (\hat{G}_1)^{31} \\
(\hat{G}_2)^{12} & (\hat{G}_2)^{23} & (\hat{G}_2)^{31} \\
(\hat{G}_3)^{12} & (\hat{G}_3)^{23} & (\hat{G}_3)^{31} \\
\end{array}\right)
\end{displaymath}

\begin{displaymath}
\boldsymbol{G}_3=
\left(\begin{array}{ccc}
(\hat{G}_1)^{11} & (\hat{G}_1)^{22} & (\hat{G}_1)^{33} \\
(\hat{G}_2)^{11} & (\hat{G}_2)^{22} & (\hat{G}_2)^{33} \\
(\hat{G}_3)^{11} & (\hat{G}_3)^{22} & (\hat{G}_3)^{33} \\
\end{array}\right)
\end{displaymath}

\noindent
Correspond to the diffeomorphisms, and

\begin{displaymath}
\boldsymbol{H}_1=
\left(\begin{array}{ccc}
0 & 0 & 0 \\
\hat{\partial}^{21} & \hat{\partial}^{32} & \hat{\partial}^{13} \\
\hat{\Delta}^{21} & \hat{\Delta}^{32} & \hat{\Delta}^{13} \\
\end{array}\right);~~
\boldsymbol{H}_2=
\left(\begin{array}{ccc}
0 & 0 & 0 \\
\hat{\partial}^{12} & \hat{\partial}^{23} & \hat{\partial}^{31} \\
\hat{\Delta}^{12} & \hat{\Delta}^{23} & \hat{\Delta}^{31} \\
\end{array}\right)\end{displaymath}

\begin{displaymath}
\boldsymbol{H}_3=
\left(\begin{array}{ccc}
1 & 1 & 1 \\
\hat{\partial}^{11} & \hat{\partial}^{22} & \hat{\partial}^{33} \\
\hat{\Delta}^{11} & \hat{\Delta}^{22} & \hat{\Delta}^{33} \\
\end{array}\right)
\end{displaymath}

\noindent
the Hamiltonian contribution.  As a practical method to compute the inverse of the original nine by nine matrix we first `normalize' it by multiplying through by the inverse of the block three by three matrix forming the diagonal, given by

\begin{displaymath}
D_{ab}^{cd}=
\left(\begin{array}{ccc}
\boldsymbol{D}_1 & 0 & 0\\
0 & \boldsymbol{G}_2 & 0\\
0 & 0 & \boldsymbol{H}_3\\
\end{array}\right).
\end{displaymath}

\noindent
The significance of the matrix $D_{ab}^{cd}$ is that it forms a reducible representation of the matter sources, whereas the original matrix $O_{ab}^{cd}$ is irreducible.  Then we compose the inverse of the diagonal elements with the off-diagonal ones.  It might be more convenient to compute the inverse of a set of three by three rather than a nine by nine matrix.  The composed matrix is given by

\begin{displaymath}
U_{ab}^{cd}=
\left(\begin{array}{ccc}
I_{(3)} & (\boldsymbol{D}_1)^{-1}\boldsymbol{D}_2 & (\boldsymbol{D}_1)^{-1}\boldsymbol{D}_3\\
(\boldsymbol{G}_2)^{-1}\boldsymbol{G}_1 & I_{(3)} & (\boldsymbol{G}_2)^{-1}\boldsymbol{G}_3\\
(\boldsymbol{H}_3)^{-1}\boldsymbol{H}_1 & (\boldsymbol{H}_3)^{-1}\boldsymbol{H}_2 & I_{(3)}\\
\end{array}\right),
\end{displaymath}

\noindent
where $I_{(3)}$ denotes the unit three by three matrix.  In the case when the operators comprising the block submatrices do in fact commute, then the inverse can then be found using the Schur decomposition 

\begin{displaymath}
\left(\begin{array}{cc}
A & B \\
C & D\\
\end{array}\right)^{-1}=
\left(\begin{array}{cc}
I_{(6)} & 0 \\
-D^{-1}C & I_{(3)}\\
\end{array}\right)
\left(\begin{array}{cc}
(A-BC)^{-1} & 0 \\
0 & D^{-1}\\
\end{array}\right)
\left(\begin{array}{cc}
I_{(6)} & -BD^{-1} \\
0 & I_{(3)}\\
\end{array}\right)
\end{displaymath}

\noindent
by choosing

\begin{displaymath}
A=
\left(\begin{array}{cc}
I_{(3)} & (\boldsymbol{D}_1)^{-1}\boldsymbol{D}_2 \\
(\boldsymbol{G}_2)^{-1}\boldsymbol{G}_1 & I_{(3)}\\
\end{array}\right);~~B=
\left(\begin{array}{c}
(\boldsymbol{D}_1)^{-1}\boldsymbol{D}_3 \\
(\boldsymbol{G}_2)^{-1}\boldsymbol{G}_3\\
\end{array}\right)
\end{displaymath}

\begin{displaymath}
C=
\left(\begin{array}{cc}
(\boldsymbol{H}_3)^{-1}\boldsymbol{H}_1 & (\boldsymbol{H}_3)^{-1}\boldsymbol{H}_2 \\
\end{array}\right);~~D=\boldsymbol{I}
\end{displaymath}

\noindent
and then reapplying the Schur decomposition to the six by six matrix of integro-differential operators $(A-BC)^{-1}$.  The premise is that one can invert the corresponding matrix differential operators by applying the appropriate boundary conditions, to obtain a finite solution.  In the general case when these operator-valued matrices do not commute, special techniques must be developed to take into account the ordering.

\section{The introduction of matter fields}

The origin of the matter terms can be made more clear when one views them as back-reactions on spacetime.  In a universe devoid of gravitational interactions the matter fields would be described by an action $S_{matter}[\phi,\partial\phi]$ defined on Minkowski spacetime with the following identifications\footnote{Assuming that Poincare symmetry is a basic symmetry of nature in this limit.} 

\begin{eqnarray}
\label{GLOBAL}
S_{matter}=\int{dt}L(t)=\int{dt}\int_{\Sigma}d^3{x}L(\phi,\partial\phi);~~\pi_{\alpha}=\Bigl({{\delta{L}} \over {\delta\dot{\phi}^{\alpha}}}\Bigr).
\end{eqnarray}

\noindent
The relevant symmetries on this spacetime are invariance under global spacetime translations and global Lorentz symmetry.  Note that due to the absence of gauge fields $A^a_i$ for the matter fields to couple to,\footnote{Due to the absence of gravity in this limit.} there are no local symmetries in the pure matter theory.\par
\indent
By the Noether theorem, for every global symmetry of a theory there exists a locally conserved current, the time components of which correspond to conserved charges.  The analogous local symmetry would be a semidirect product of $SU(2)_{-}$ gauge transformations and diffeomorphisms, however since there is no gravity the corresponding global symmetry is a semidirect product of Lorentz invariance with spacetime translations.  The Noether charges corresponding to these symmetries are the local 4-momentum and Lorentz charge integrated over all space

\begin{eqnarray}
\label{GLOBALL1}
T_{00}=P_0=\int_{\Sigma}d^3{x}\bigl(\pi_{\alpha}(x)\dot{\phi}^{\alpha}(x)-L\bigr);\nonumber\\
~~T_{0i}=P_i=-\int_{\Sigma}d^3{x}\pi_{\alpha}(x)\partial_i\phi^{\alpha}(x);\nonumber\\
Q_a=\int_{\Sigma}d^3{x}\pi_{\alpha}(x)(T_a)^{\alpha}_{\beta}\phi^{\beta}(x).
\end{eqnarray}

\noindent
Equation (\ref{GLOBALL1}) corresponds to a total of seven global symmetries.  Note that in the gravitationally coupled theory it is precisely the local versions of these charges which consistute the matter contributions to the Hamiltonian, diffeomorphism, and Gauss' law constraints respectively 
\cite{ASH},\cite{ASH1},\cite{ASH2},\cite{ASH3}.  By our new method the local form of these conserved charges constitute one of the inputs into (\ref{FORM}) in order to produce the corresponding equations for the matter-coupled theory.  These quantities must be consistent with a proper semiclassical limit of the gravity-coupled theory.  In the absence of gravity we have 
$\pi_{\alpha}(A^a_i(x),\phi^{\beta}(x))=f_{\alpha}(\phi^{\beta}(x))$, whereupon the mixed partials condition becomes trivialized.  The inputs from the gravity-free sector into the interaction sector can directly be derived from $f_{\alpha}$, and include

\begin{eqnarray}
\label{GLOBAL1}
\widetilde{\tau}_{00}={{T_{00}[f]} \over {\hbox{det}B}};\nonumber\\
~~\widetilde{\tau}_{0d}={{B^i_d{T}_{0i}} \over {\hbox{det}B}}=(\hbox{det}B)^{-1}B^i_df_{\alpha}(\phi(x))D_i\phi^{\alpha}(x);\nonumber\\
Q_a=\lambda{f}_{\alpha}(\phi(x))(T_a)^{\alpha}_{\beta}\phi^{\beta}(x).
\end{eqnarray}

\noindent
Equation (\ref{GLOBAL1}) is simply equation (\ref{GLOBALL1}) with the momenta $\pi_{\alpha}$ replaced by their functional boundary condition $f_{\alpha}$ in the mixed partials condition, projected completely into $SU(2)_{-}$ and densitized by $\hbox{det}B$ with weight $-1$.  This implies that the Ashtekar magnetic field $B^i_a$ must be nondegenerate, which can also be seen from the semiclassical contribution to the Hamiltonian and diffeomorphism constraints

\begin{eqnarray}
\label{HAMMMM}
\hbox{det}B\bigl(Var\Psi+\Lambda\hbox{det}\Psi\bigr)+GT_{00}=0;~~\hbox{det}B\epsilon_{dae}\Psi_{ae}=B^i_dT_{0i}.
\end{eqnarray}

\noindent
One result of the nondegeneracy of $B^i_a$ is that it allows for the possibility of a direct link to Einstein's general relativity to be maintained throughout the quantization procedure.  Note that the first order formulations of general relativity are generalizations of the second order formulation which allow for the existence of a degenerate metric.  A degenerate magnetic field in the CDJ representation signifies a degenerate metric which allows for the possibility of topology change \cite{DEGENERATE}.\footnote{Recall that in order to recover the second-order formalism of general relativity from the first-order formalism, it is a necessary condition that the tetrad be nondegenerate.  Therefore, one may conclude that the generalized Kodama states, due to the presence of matter fields, do not allow for changes in topology.  However, for each given topological sector, one can still compute the associated $\Psi_{GKod}$.}  The existence of matter fields in coupled to gravity in Ashtekar variables implies this nondegeneracy through the CDJ Ansatz $\widetilde{\sigma}^i_a=\Psi_{ae}B^i_e$.  Note, also, that in order for general relativity to be consistently coupled to typical matter fields the contravariant metric is needed, which as well requires nondegeneracy.\par
\indent
The function $f_{\alpha}$ can be identified with the one-point 1PI vertex of the effective action for the matter theory, quantized in the absence of gravity.

\begin{eqnarray}
\label{IREE}
f_{\alpha}[\vec{\phi}]=
{{\delta\Gamma_{eff}[\vec{\phi}]} \over {\delta\phi^{\alpha}(\boldsymbol{x},T)}}
\end{eqnarray}

\noindent
The matter source vector can be written in the following general form

\begin{eqnarray}
\label{FORM3}
Q_{ae}=\sum_{d=1}^{3}\widetilde{\tau}_{0d}\epsilon_{[dae]}
+\sum_{d=1}^{3}Q_{d}\epsilon_{(dae)}+(\Lambda/12)\widetilde{\tau}_{00}\delta_{g1}\delta_{h1}
+q\delta_{g2}\delta_{h2}+\mu\delta_{g3}\delta_{h3}.
\end{eqnarray}

\noindent
As a review of \cite{EYOFULL}, $\widetilde{\tau}_{0d}$ are the space-time components of the total energy momentum tensor of the matter fields projected into $SU(2)$ 
while $\widetilde{\tau}_{00}$ is the time-time component, both densitized by $\vert{B}\vert$ with weight $-1$.  The quantities $Q_a$ are the 
matter $SU(2)_{-}$ charges, which are undensitized because they already take their values in $SU(2)_{-}$.\par
\indent  
The quantity $q$ corresponds to the functional divergence of the matter momentum which is a quantum effect, due to matter Hamiltonians quadratic in momenta, occuring at the first order in singularity 
$\hbar{G}\delta^{(3)}(0)$.  The physical interpretation of the functional divergence term is simply the trace of the mass squared matrix of the quantized gravity-free theory of matter.  This matrix takes into account all the matter fields of the theory and can be derived from the effective potential

\begin{eqnarray}
\label{FUNCDIV}
q=-{i \over {4G}}{{\partial{f}^{\alpha}} \over {\partial\phi^{\alpha}}}\sim-{i \over {4G}}
\Bigl({{\partial^2V_{eff}[\vec{\phi}]} \over {\partial\phi^{\alpha}\partial\phi_{\alpha}}}\Bigr).
\end{eqnarray}

\noindent
Just as the matter momentum $H_i$ acts as a source for the antisymmetric part of $\epsilon_{ae}$, $q$ in (\ref{FUNCDIV}) acts as a source for functional divergence of $\epsilon_{ae}$ due to quantum fluctuations first order in singularity.\par  
\indent
The quantity $\mu$ is an effect occuring at the second order of singularity $(\hbar{G}\delta^{(3)}(0))^2$, which usually encodes the ordering ambiguities in the operators comprising the Hamiltonian constraint.  The quantity $\mu$ acts as a source for the functional Laplacian of 
$\epsilon_{ae}$.  Upon coupling to the Ashtekar variables, the matter input to the constraints gets rearranged into the following form

\begin{displaymath}
\left(\begin{array}{ccc}
Q^{\prime}_{11} & Q^{\prime}_{12} & Q^{\prime}_{13}\\
Q^{\prime}_{21} & Q^{\prime}_{22} & Q^{\prime}_{33}\\
Q^{\prime}_{31} & Q^{\prime}_{32} & Q^{\prime}_{33}\\
\end{array}\right)=
\left(\begin{array}{ccc}
(\Lambda/12)\widetilde{\tau}_{00} & Q_3 & \widetilde{\tau}_{02}\\
\widetilde{\tau}_{03} & \widetilde{q} & Q_1\\
Q_2 & \widetilde{\tau}_{01} & \widetilde{\mu}\\
\end{array}\right)
\end{displaymath} 

\noindent
The corresponding set of constraints for the matter-coupled theory can be written in the concise form

\begin{eqnarray}
\label{FORM4}
\Sigma^{ghae}\epsilon_{ae}+\Lambda\Sigma^{ghaebf}\epsilon_{ae}\epsilon_{bf}
+\Lambda^2{E}^{ghabcdef}\epsilon_{ad}\epsilon_{be}\epsilon_{cf}=G{Q^{\prime}}^{gh}.
\end{eqnarray}

\noindent
where the operator $\Sigma^{ghae}$ is in general a nine by nine matrix of integro-differential operators, which can be written in the general compact form

\begin{eqnarray}
\label{FORM5}
\Sigma^{ghae}=\sum_{d=1}^{3}\sigma_{[aed]}\epsilon_{dgh}
+\sum_{d=1}^{3}\sigma^{(ae)}_{d}\epsilon_{(aed)}+\sigma^{ae}\delta_{g1}\delta_{h1}+\nabla^{ae}\delta_{g2}\delta_{h2}+\Delta^{ae}\delta_{g3}\delta_{h3}
\end{eqnarray}

\noindent
and the following quantities are defined

\begin{eqnarray}
\label{FORM6}
\Sigma^{ghabef}={1 \over 6}\delta_{g1}\delta_{h1}\bigl(\delta_{ae}\delta_{bf}-\delta_{af}\delta_{be}+\sigma_{abef}\bigr)
+{1 \over 8}\delta_{g2}\delta_{h2}\nabla^{eb}_{af}.
\end{eqnarray}

\noindent
where $\sigma^{ae}$ and $\sigma_{aebf}$ are operators containing dependence on the spatial part of the energy momentum tensor $T_{ij}$ of matter, projected 
into $SU(2)_{-}$ and densitized by $\vert{B}\vert$ with weight $-1$.\par
\indent
The matter contribution to the constraints gets injected into three slots in (\ref{FORM}) in the matter-coupled theory: (i) on the right hand side to act as a Noetherian source, 
(ii) to the linear term to correct the $SU(2)_{-}$ DeSitter metric, (iii) To the quadratic term to correct the contribution to the variance.  The first slot corresponds to the matrix $Q^{\prime}_{ab}$ shown above.  The second and third slots incorporate the source terms for the space-space part of the Einstein's equations $\widetilde{\tau}_{ae}$, or the spatial part of the energy momentum tensor, also a Noether charge, projected into $SU(2)$ and densitized.\par
\indent
The CDJ deviation matrix appears in the form

\begin{displaymath}
\epsilon_{ae}=
\left(\begin{array}{ccc}
\epsilon_{11} & \epsilon_{12} & \epsilon_{13}\\
\epsilon_{21} & \epsilon_{22} & \epsilon_{33}\\
\epsilon_{31} & \epsilon_{32} & \epsilon_{33}\\
\end{array}\right)
\end{displaymath} 

\noindent
as dictated by the constraints (\ref{FORM4}).  However, the elements of this matrix do not match up to the corresponding elements of the source $Q^{\prime}_{ae}$.  This due to the fact that the kinetic operator $\Sigma^{aebf}$ is not a diagonal matrix.  In order to correctly assess the relation of Einstein's general relativity to Ashtekar variables in the semiclassical limit, one must express $\epsilon_{ae}$ in terms of a more convenient set of variables which explicitly indicate what physical quantities the Noetherian charges correlate to.  In analogy to the case of pure gravity, one does not invert the matrix $\Sigma^{aebf}$ directly, but instead transforms it into a (block) diagonal matrix which gets inverted.  The significance of the block diagonal matrix $D_{ab}^{cd}$ is that it forms a reducible representation of the Noetherian source 9-vector.  Left-multiplying this matrix into (\ref{FORM4}) yields

\begin{eqnarray}
\label{FORM7}
\psi_{mn}=R_{mn}^{gh}\epsilon_{gh}=G(D^{-1})_{mngh}{Q^{\prime}}^{gh}\nonumber\\-\Lambda(D^{-1})_{mngh}\Sigma^{ghaebf}\epsilon_{ae}\epsilon_{bf}
-\Lambda^2(D^{-1})_{mngh}{E}^{ghabcdef}\epsilon_{ad}\epsilon_{be}\epsilon_{cf}.
\end{eqnarray}

\noindent
where $R_{mn}^{gh}$ is a generalized rotation matrix in this abstract group of integro-differential transformations, with $\psi_{mn}$ being the `rotated' 
version of $\epsilon_{ae}$, given by

\begin{eqnarray}
\label{FORM8}
R_{mn}^{gh}=(D^{-1})_{mnef}\Sigma^{efgh};~~\epsilon_{gh}=(R^{-1})_{gh}^{mn}\psi_{mn}.
\end{eqnarray}

\noindent
Note that the matrices $R_{mn}^{gh}$ and $(D^{-1})_{mnef}$ are model-specific, corresponding to the matter content of the theory.  The constraints can then be written in the form

\begin{eqnarray}
\label{FORM9}
\psi_{mn}=G(D^{-1})_{mngh}{Q^{\prime}}^{gh}
-\Lambda(D^{-1})_{mngh}\Sigma^{ghaebf}(R^{-1})_{ae}^{a^{\prime}e^{\prime}}(R^{-1})_{bf}^{b^{\prime}f^{\prime}}    
\psi_{a^{\prime}e^{\prime}}\psi_{b^{\prime}f^{\prime}}\nonumber\\
-\Lambda^2(D^{-1})_{mngh}{E}^{ghabcdef}(R^{-1})_{ad}^{a^{\prime}d^{\prime}}(R^{-1})_{be}^{b^{\prime}e^{\prime}}(R^{-1})_{cf}^{c^{\prime}f^{\prime}}
\psi_{a^{\prime}d^{\prime}}\psi_{b^{\prime}e^{\prime}}\psi_{c^{\prime}f^{\prime}}.
\end{eqnarray}

\noindent
There are a few significant differences from (\ref{FORM9}) to its matter-free (\ref{FORMM}).  (i) First, there is a inhomogeneous source term $Q^{\prime}_{gh}$ which is suppressed by a factor of $G$ on the right hand side of (\ref{FORM9}).  It is consistent with the Einstein metric equations of general relativity for the matter source terms to be suppressed by a factor of $G$ relative to the matter-free terms.  When one interprets $(D^{-1})_{mngh}$ as the (block diagonal) propagator, one sees that the matter terms act as a direct source for the corresponding elements of $\psi_{ae}$, the rotated form of $\epsilon_{ae}$.  Hence there is a direct correspondence at the linearized level between the matrix elements

\begin{displaymath}
\psi_{ae}=
\left(\begin{array}{ccc}
\psi_{11} & \psi_{12} & \psi_{13}\\
\psi_{21} & \psi_{22} & \psi_{33}\\
\psi_{31} & \psi_{32} & \psi_{33}\\
\end{array}\right)\sim
\left(\begin{array}{ccc}
Q^{\prime}_{11} & Q^{\prime}_{12} & Q^{\prime}_{13}\\
Q^{\prime}_{21} & Q^{\prime}_{22} & Q^{\prime}_{33}\\
Q^{\prime}_{31} & Q^{\prime}_{32} & Q^{\prime}_{33}\\
\end{array}\right)
\end{displaymath} 

\noindent
which makes it easier to keep track of the progression of the source.  (ii) One creates a tree diagram from the recursion relation by making the following identifications.  Draw a straight line for each matter-modified propagator $(D^{-1})_{mngh}$ and an $X$ with a corresponding factor of $G$ for each source element $Q^{\prime}_{ae}$.  Label the two types of matter vertices (bivalent and trivalent respectively) with 
factors of $\Lambda$ and $\Lambda^2$, along with two or three circles to correspond to the rotation matrix $(R^{-1})_{gh}^{mn}$.  (iii)  One draws the corresponding network and discovers that the there are an infinite number of terms.  However, the lines of the network all originate from matter sources $Q^{\prime}_{ae}$ (with associated factor of $G$) and connect to the two types of vertices in all possible ways, with the associated powers of $\Lambda$.  Whereas all lines of the network in the matter-free case have no beginning and have no end other than on vertices, when coupled to matter the lines terminate (or originate) on these sources giving each leg of the network a finite length.  The first few terms of the series are given by

\begin{eqnarray}
\label{FORM10}
\psi_{cd}={1 \over \Lambda}\biggl[(G\Lambda)(D^{-1})_{cd}^{ab}Q^{\prime}_{ab}
+(G\Lambda)^2(D^{-1})_{cd}^{ab}\Sigma_{ab}^{cdef}
(R^{-1})_{cd}^{c^{\prime}d^{\prime}}(R^{-1})_{ef}^{e^{\prime}f^{\prime}}Q^{\prime}_{c^{\prime}d^{\prime}}Q^{\prime}_{e^{\prime}f^{\prime}}\nonumber\\
+(G\Lambda)^3(D^{-1})_{cd}^{ab}E^{cdefgh}_{ab}(R^{-1})_{cd}^{c^{\prime}d^{\prime}}(R^{-1})_{ef}^{e^{\prime}f^{\prime}}
(R^{-1})_{gh}^{g^{\prime}h^{\prime}}Q^{\prime}_{c^{\prime}d^{\prime}}Q^{\prime}_{e^{\prime}f^{\prime}}Q^{\prime}_{g^{\prime}h^{\prime}}+...\biggr]
\end{eqnarray}

\noindent
Since the only two types of vertices allowed are the bivalent and the trivalent vertices, then the number and types of allowed terms for any given order in the pertubative expansion is limited by network topology.    This, combined with the suppresion of each occurence of the source by a factor of $G\Lambda$ hopefully ensures rapid convergence of the series (at a rate of the dimensionless coupling constant $(G\Lambda)^N$).  The diagrams that accompany these factors are model specific.  (iv) In the case of pure gravity, the solution was zero due to the factors of $\Lambda$ being carried through an infinite number of iterations.  In the matter-coupled case there is more flexibility in the solution.  Since one inputs the observable semiclassical limit for $Q^{\prime}_{ae}$, the result of the gravitational sector should be such as to extrapolate this limit, through a rapidly proliferating sequence of operations and terms, to something that hopefully should be clearly discernible by experiment.  All models have the same network, albeit with different vertices and propagators.  Thus the effects of a model should be distinguishable.\par
\indent

\section{The quantum theory of deviations from DeSitter spacetime}

\noindent
We have argued in \cite{EYOPATH} the equivalence between the path integration and canonical methods to quantization by way of the generalized Kodama states $\Psi_{GKod}$, by exploiting the preservation of the functional form of the constraints on each spatial hypersurface $\Sigma_t$ for each $t_0\leq{t}\leq{T}$.  This correspondence implies that the path integral quantization of gravity in Ashtekar variables converges and is finite, assuming that the state itself is finite.  In configuration space this corresponds to the quantum field theory of fields on spacetime.  The quantum theory exhibits a fibre bundle structure with the spacetime 
manifold $M=\Sigma\otimes{R}$ serving as the base space and the fields $(A^a_i(x),\phi^{\alpha}(x))$ in $\Gamma$ serving as the fibre, with the structural 
group $SU(2)_{-}\otimes{Diff}_M$.  We are now going to study a new type of structure within the existing structure.\par
\indent
Prior to the solution of the constraints there was no restriction on $\epsilon_{ae}$.  Once solved, this variable acquires dependence upon the fields 
$(A^a_i(x),\phi^{\alpha}(x))$ comprising $\Gamma$ at each spatial point in $\Sigma$.  Consider the fibre bundle structure $\epsilon_{ae}$ in which the configuration variables form the base space at each point in $\Sigma$ with $Q^{\prime}_{ae}$ comprising the fibre.  The structural group is given by the matrix $R^{ab}_{cd}$, which as we recall from the diagrammatic expansion `rotates' the disturbance from the source after propagation into the correct orientation to connect to a either a trivalent or a tetravalent vertex.\par
\indent
We will now study the quantum field theory on this new bundle structure, in direct analogy to the second-quantization of field theory.  In a certain sense this is a third-quantization since the dynamical variables $(A^a_i,\phi^{\alpha})$ that were previously second-quantized have essentially become labels.  Prior to quantization of the theory on this new space one must first identify an action which leads to the governing equations, in this case the constraints $C_{ab}=0$, of motion.

\subsection{Pure gravity case}

\noindent
In guessing the starting action corresponding to the equations of the motion $C_{ab}=0$ one must look for the simplest action that (i) reproduces the quantized constraints of general relativity, and (ii) correctly implements the Feynman rules.  The simplest action producing the constraints (\ref{FORM}) is given by

\begin{eqnarray}
\label{QUANTUM}
\Gamma^{\prime}(\lambda,\epsilon)=\int{D\mu}\bigl(\lambda_{gh}O^{ghae}\epsilon_{ae}+\Lambda\lambda_{gh}{I}^{ghaebf}\epsilon_{ae}\epsilon_{bf}
+\Lambda^2\lambda_{gh}{E}^{ghabcdef}\epsilon_{ad}\epsilon_{be}\epsilon_{cf}\bigr).
\end{eqnarray}

\noindent
where $\lambda_{ab}$ is an auxilliary field and where the measure $D\mu$ is over the functional space of connections at each point, given by

\begin{eqnarray}
\label{QUANTUM1}
D\mu=\prod_{a,i}dA^a_i(x).
\end{eqnarray}
  
Equation (\ref{QUANTUM}) is precisely the term occuring in \cite{EYOPATH} in the noboundary approach, when the wavefunction of the universe is taken off-shell.  

\noindent
The equations of motion stemming from (\ref{QUANTUM}) are given by

\begin{eqnarray}
\label{QUANTUM2}
\Bigl({{\delta\Gamma^{\prime}} \over {\delta\lambda_{ae}}}\Bigr)=C_{ae}=0.
\end{eqnarray}

\noindent
Although $\lambda_{ab}$ plays the role of an auxilliary field to implement the constraints, we will show that it has a physical interpretation relevant to the dynamics.  In the path integral quantization of the action one must evaluate

\begin{eqnarray}
\label{QUANTUM3}
Z=\int{D\mu[\psi]}{D\mu[\lambda]}e^{i\Gamma^{\prime}(\psi,\lambda)}
\end{eqnarray}

\noindent
and its relevant correlation functions, where the path integration measure is given on the functional space of fields $A^a_i$ for each point $x$ by

\begin{eqnarray}
\label{QUANTUM4}
\int{D}\mu[\psi]{D}\mu[\lambda]=\prod_{a,e,b,f}\prod_{A^a_i(x)}d\psi_{ae}[A^a_i(x)]d\lambda_{bf}[A^a_i(x)].
\end{eqnarray}

\noindent
So the path integral measure is infinite dimensional in that it ranges over all values of $\psi_{ab}$ and $\lambda_{ab}$ for a particular function $A^a_i(x)$ at the point $x$ for all functions.  Since there is an infinite number of functions, the measure is infinite dimensional.  When one considers that the path integral can be performed identically for each point $x$ in $M$, then one sees that there is a separate infinite dimensionality, apart from that on the space of functions, due to the infinite number of points in $M$.  We are evaluating the quantum field theory of fluctuations from DeSitter spacetime at one particular point.  Since the constraints and the path integral have the same functional form irrespective of $x$, the analysis extends to all points $x$ of $M$.  It resembles minisuperspace, but is in fact still the full theory.\par
\indent
Now that we have specified the arena for the quantization, we will perform (\ref{QUANTUM3}) for the matter-free case in order to obtain an interpretation for the physics involved.  The starting action (\ref{QUANTUM}) resembles a kind of distorted $\phi^4$ theory on the space of functions.  We will first perform the free part of the path integral, which requires the introduction of current sources.  We will need one source $j^{ab}$ for $\psi_{ab}$ and another source $J^{ab}$ 
for $\lambda_{ab}$.  These sources are not to be interpreted as matter currents, and are provided simply to enable evaulation of the perturbative series.  So the free part (\ref{QUANTUM3}) is given by

\begin{eqnarray}
\label{QUANTUM5}
Z_{0}(J,j)=\int{D\mu[\psi]}{D\mu[\lambda]}e^{(i\lambda\cdot{D}\psi+\lambda\cdot{J}+\psi\cdot{j})},
\end{eqnarray}

\noindent
In (\ref{QUANTUM5}) the matrix $D$ is a matrix of functional integro-differential operators on the space of functions $A$, given by

\begin{displaymath}
D_{ab}^{cd}=
\left(\begin{array}{ccc}
\boldsymbol{D}_1 & 0 & 0\\
0 & \boldsymbol{G}_2 & 0\\
0 & 0 & \boldsymbol{H}_3\\
\end{array}\right).
\end{displaymath}

\noindent
The transformation $\psi_{ab}=R_{ab}^{cd}\epsilon_{cd}$ is now clear.  It rotates the CDJ deviation matrix into a frame in which the kinetic operator can be separated into its appropriate subspaces.  We also in (\ref{QUANTUM5}) have made the identifications

\begin{eqnarray}
\label{QUANTUM6}
\lambda\cdot{J}=\int{d\mu[A]}{d\mu[A^{\prime}}]\prod\delta(A(x)-A^{\prime}(x))\lambda_{ab}[A(x)]J^{ab}[A^{\prime}(x)];\nonumber\\
\lambda\cdot{j}=\int{d\mu[A]}{d\mu[A^{\prime}}]\prod\delta(A(x)-A^{\prime}(x))\psi_{ab}[A(x)]j^{ab}[A^{\prime}(x)]\nonumber\\
\end{eqnarray}

\noindent
for the sources, and

\begin{eqnarray}
\lambda\cdot{D}\psi=\int{d\mu[A]}{d\mu[A^{\prime}]}\lambda_{ab}[A(x)](\boldsymbol{H}_3)^{ab}_{cd}[A(x),A^{\prime}(x)]\psi_{ab}[A^{\prime}(x)]\nonumber\\
+\int{Dx^{\prime}}{Dx}\lambda_{ab}[A(x^{\prime})](\boldsymbol{G}_2)^{ab}_{cd}(x,x^{\prime})\psi_{ab}[A(x)]\nonumber\\
+\int{Dx^{\prime}}{Dx}\lambda_{ab}[A(x^{\prime})]
(\boldsymbol{D}_1)^{ab}_{cd}(x,x^{\prime})\psi_{ab}[A(x)]
\end{eqnarray}

\noindent
for the kinetic contribution.  We now perform the Gaussian integral

\begin{eqnarray}
\label{QUANTUM7}
Z_{0}(J,j)=\int{D\mu[\psi}]{D\mu[\lambda]}e^{i(\lambda\cdot{D}\psi-i\lambda\cdot{J}-i\psi\cdot{j})}
=\int{D\mu[\psi]}\prod_{A}\delta(D\psi-iJ)e^{\psi\cdot{j}}\nonumber\\
=\int{D\mu[\psi]}\prod_{A}(\hbox{Det}^{-1}D)\delta(\psi-iD^{-1}J)e^{\psi\cdot{j}}
=(\hbox{Det}D)^{-1}e^{iJ\cdot{D^{-1}j}}.
\end{eqnarray}

\noindent
The operator $D^{-1}$, given by

\begin{displaymath}
(D^{-1})_{ab}^{cd}=
\left(\begin{array}{ccc}
(\boldsymbol{D}_1)^{-1} & 0 & 0\\
0 & (\boldsymbol{G}_2)^{-1} & 0\\
0 & 0 & (\boldsymbol{H}_3)^{-1}\\
\end{array}\right).
\end{displaymath}

\noindent
acts as a propagator which governs the interaction between sources.  In the case of the Gauss' law constraint one can use the result from \cite{EYOGAUSS} that

\begin{displaymath}
(\boldsymbol{G}_2)^{-1}=
\left(\begin{array}{ccc}
{\bigl({\partial \over {\partial{t^2}}}+C_1^{12}\bigr)} & {C_1^{23}} & {C_1^{31}}\\
{C_2^{12}} & {\bigl({\partial \over {\partial{t^3}}}+C_3^{23}\bigr)} & {C_2^{31}}\\
{C_3^{12}} & {C_3^{23}} & {\bigl({\partial \over {\partial{t^1}}}+C_1^{31}\bigr)}\\
\end{array}\right)^{-1}
\end{displaymath}

\noindent
in which case the propagation takes place between two spatial points $(t^e,{t^{\prime}}^a)$ in $\Sigma$ for a fixed function $A=A^a_i$.  The corresponding kernel is nonlocal and signifies an instantaneous interaction of sources separated by a distance.  For the Hamiltonian constraint one must invert the matrix of functional differential operators 

\begin{displaymath}
(\boldsymbol{H}_3)^{-1}=
\left(\begin{array}{ccc}
1 & 1 & 1\\
\partial^{11} & \partial^{22} & \partial^{33}\\
\Delta^{11} & \Delta^{22} & \Delta^{33}\\
\end{array}\right).
\end{displaymath}

\subsection{Incorporation of the interactions in the matter-free case}

\noindent
Now that we have found the Gaussian part of the path integral we can now make the identification

\begin{eqnarray}
\label{QUANTUM8}
Z(J,j)= e^{\Lambda\int{D\mu_A}{V}_2(\delta/\delta{J},\delta/\delta{j})}
e^{\Lambda^2\int{D\mu_A}{V}_3(\delta/\delta{J},\delta/\delta{j})}e^{iJ\cdot{D^{-1}j}}
\end{eqnarray}

\noindent
where we have defined

\begin{eqnarray}
\label{QUANTUM9}
V_2={I}^{ghaebf}{\delta \over {\delta{J}^{gh}}}{\delta \over {\delta{j}^{ae}}}{\delta \over {\delta{j}^{bf}}};\nonumber\\
V_3={E}^{ghabcdef}{\delta \over {\delta{J}^{gh}}}{\delta \over {\delta{j}^{ad}}}{\delta \over {\delta{j}^{be}}}{\delta \over {\delta{j}^{cf}}}
\end{eqnarray}

\noindent
To get an idea for the result, let us expand in a triple series.  Schematically, suppressing indices to avoid cluttering up the notation,

\begin{eqnarray}
\label{QUANTUM10}
Z(J,j)=\hbox{Det}^{-1}D\sum_{lmn}{{\Lambda^{l+2m}} \over {l!m!n!}}(Symmetry~factors)\Bigl[{{\delta^3} \over {\delta{J}\delta{j}\delta{j}}}\Bigr]^l
\Bigl[{{\delta^4} \over {\delta{J}\delta{j}\delta{j}\delta{j}}}\Bigr]^m(iJ\cdot{D^{-1}j})^n
\end{eqnarray}

\noindent
Upon performing all operations in (\ref{QUANTUM10}) one must set the sources $(J,j)$ to zero to obtain $Z_0$ and collect the remaining nontrivial terms.  But it is easy to see from balancing powers of $J$ 
that $l+m=n$, and that $2l+3m=n$ due to balancing powers of $j$.  This imples that $l+2m=0$, or $l=m=n=0$ since there are nonnegative integers.  So the partition function for the pure gravity case, which correspond 
to $\Psi_{Kod}$ produces 

\begin{eqnarray}
\label{QUANTUM11}
Z(0,0)=\Bigl[(\hbox{Det}\boldsymbol{D}_1)(\hbox{Det}\boldsymbol{G}_2)(\hbox{Det}\boldsymbol{H}_3)\Bigr]^{-1}.
\end{eqnarray}

One can re-interpret (\ref{FORM5}) in terms of a `renormalized' 
charge $q^{gh}$ by making the identification

\begin{eqnarray}
\label{FORM7}
\Sigma^{ghae}\epsilon_{ae}=Gq^{gh}\longrightarrow\epsilon_{ae}=G\Sigma^{-1}_{aegh}q^{gh}.
\end{eqnarray}

\noindent
This has the interpretation of charge required to produce a solution 
to $\epsilon_{ae}$ at the linearized level.  Upon substitution back 
into (\ref{FORM4}) this leads to

\begin{eqnarray}
\label{FORM8}
{Q^{\prime}}^{gh}=q^{gh}+G\Lambda(I+\Sigma)^{ghaebf}
(O+\Sigma)^{-1}_{aea^{\prime}e^{\prime}}(O+\Sigma)^{-1}_{bfb^{\prime}f^{\prime}}q^{a^{\prime}e^{\prime}}q^{b^{\prime}f^{\prime}} \nonumber\\
+(G\Lambda)^2{E}^{ghabcdef}(O+\Sigma)^{-1}_{ada^{\prime}d^{\prime}}(O+\Sigma)^{-1}_{beb^{\prime}e^{\prime}}(O+\Sigma)^{-1}_{cfc^{\prime}f^{\prime}}    
q^{a^{\prime}d^{\prime}}q^{b^{\prime}e^{\prime}}q^{c^{\prime}f^{\prime}}.
\end{eqnarray}

\noindent
We will develop the perturbative aspects in greater detail in \cite{EYOCONS}, in which we treat the gravity and matter fields as one unified field.  To handle the more general case of the matter coupled theory, we make the replacements $I+\Sigma\rightarrow\Sigma$, $D^{-1}\rightarrow(D+\sigma)^{-1}$, and we do not set the currents to zero except for $J$, the source for $\lambda$.  The perturbative expansion then reads

\begin{eqnarray}
\label{MATTTEER}
Z(J,j)=\hbox{Det}^{-1}(D+\sigma)\sum_{l,m,n}{{\Lambda^{l+2m}} \over {l!m!n!}}
\prod_{k=1}^l\Sigma^{g_kh_ka_ke_kb_kf_k}
\prod_{l=1}^mE^{g_lh_la_lb_lc_ld_le_lf_l}
\prod_{q=1}^n((D+\sigma)^{-1})_{a_qe_qb_qf_q}\nonumber\\
{\delta \over {\delta{J}^{g_kh_k}}}
{\delta \over {\delta{J}^{g_lh_l}}}
{\delta \over {\delta{j}^{a_ke_k}}}
{\delta \over {\delta{j}^{b_kf_k}}}
{\delta \over {\delta{j}^{a_ld_l}}}
{\delta \over {\delta{j}^{b_le_l}}}
{\delta \over {\delta{j}^{c_lf_l}}}
(J^{a_1e_1}\dots{J}^{a_ne_n}j^{b_1f_1}\dots{j}^{b_nf_n})
\end{eqnarray}

\noindent
A simple power counting argument shows that the only nontrivial $J$ contribution survives from $l+m=n$, leaving $2l+3m$ functional derivatives to annihilate the remaining $l+m$ factors of $j$.  Therefore the only terms that survive are the $l=m=0$ term.  The net result is the following.

\begin{eqnarray}
\label{MATTTEER1}
Z(0,0)=\hbox{Det}^{-1}(O+\sigma)=\hbox{Det}^{-1}\Bigl({{\delta{C}_{ae}} \over {\delta\lambda_{bf}}} 
\Bigr)\biggl\vert_{\epsilon_{ae}=0}
=\hbox{Det}\Bigl({{\delta^2\Gamma^{\prime}} \over {\delta\lambda_{ae}\delta\epsilon_{bf}}}\Bigr)\biggl\vert_{\epsilon_{ae}=0}.
\end{eqnarray}

\noindent
Equation(\ref{MATTTEER1}) states that any quantum corrections to the classical action $\Gamma$ are limited to the one-loop term.  This seems to be in accordance with the SQC.  However, observe that there is also a contribution to the path integral representation of $\Psi_{Kod}$ from the determinant evaluated not at $\epsilon_{ae}=0$, but rather on the solution to the constraints \cite{EYOPATH}.  So we have that

\begin{eqnarray}
\label{MATTTEER2}
r\sim{\hbox{Det}\Bigl({{\delta{C}_{ae}} \over {\delta\epsilon_{bf}}}
\Bigr)\biggl\vert_{C_{ae}=0}}\hbox{Det}^{-1}\Bigl({{\delta{C}_{ae}} \over {\delta\epsilon_{bf}}} 
\Bigr)\biggl\vert_{\epsilon_{ae}=0}
\end{eqnarray}

\noindent
Evaluation of the derivative yields

\begin{eqnarray}
\label{MATTTEER3}
{{\delta{C}_{ae}} \over {\delta\epsilon_{bf}}}
=(O+\sigma)^{ghae}
+\Lambda\Sigma^{ghaebf}\epsilon_{bf}
+\Lambda^2E^{ghabcdef}\epsilon_{be}\epsilon_{cf}.
\end{eqnarray}

\noindent
Note that for the pure Kodama state $\Psi_{Kod}$, the condition $\epsilon_{ae}=0$ upon which the one-loop correction is determined coincides with the solution to the constraint.  In the presence of matter fields, there is a shift of order $G\Lambda$.  Symbolically, we have that

\begin{eqnarray}
\label{MATTTEER4}
{{\delta{C}} \over {\delta\epsilon}}=
(O+\sigma)+\Sigma(e^{G\Lambda{T}\cdot{Q}}-1)
+E(e^{G\Lambda{T}\cdot{Q}}-1)^2
\end{eqnarray}

\section{Tree networks and the transformation of pure into generalized Kodama states}

\noindent
We would like to be able to express the generalized Kodama 
state $\Psi_{GKod}$, in compact notation, as some sort of transformation of the pure Kodama state $\Psi_{Kod}$.  The transformation should have the following desirable properties: (i) The quantized gravity-free theory of matter with its associated semiclassical limit should consitute the only input into the transformation.  The rationale is that for the pure Kodama state there is a unique input 
$f_{\alpha}=0$ due to the absence of matter fields, which corresponds to the proper semiclassical limit of DeSitter spacetime.  Likewise, one hopes that in the presence of matter, the correct semiclassical limit of the gravity-free theory should lead uniquely to the generalized Kodama state. (ii) The transformation should have some kind of matrix representation acting on the nine-dimensional space of CDJ matrix elements.  The pure Kodama state should correspond to the `ground state' of the representation.  If one thinks of the CDJ matrix $\Psi_{ae}$ as a metric, then one can use the following analogy.  An infinitesimal general coordinate transformation $x^{\nu}\rightarrow\xi^{\nu}(x)$ of the Minkowski metric is given by 
$\delta_{\xi}\eta_{\mu\nu}=\xi_{\mu,\nu}+\xi_{\nu,\mu}$.  The 
metric $\eta_{\mu\nu}$ is invariant as long as $\xi^{\nu}$ is restricted to global Poincare transformations.  However, under a more general transformation the metric undergoes a change by

\begin{eqnarray}
\label{MINK}
\delta_{\xi}\eta_{\mu\nu}=\bigl(\xi^{,\beta}_{\mu}\delta^{\gamma}_{\nu}
+\xi^{,\beta}_{\nu}\delta^{\gamma}_{\mu}\bigr)\eta_{\gamma\beta}
=\xi^{,\rho}_{\sigma}(M^{\sigma}_{\rho})^{\beta\gamma}_{\mu\nu}
\eta_{\beta\gamma}
\end{eqnarray}

\noindent
where we have defined the generator of the transformation on the 10-dimensional vector space $g_{\mu\nu}$ by the quantity
$(M^{\sigma}_{\rho})^{\beta\gamma}_{\mu\nu}
=\delta^{\sigma}_{\mu}\delta^{\beta}_{\rho}\delta^{\gamma}_{\nu}
+\delta^{\sigma}_{\nu}\delta^{\beta}_{\rho}\delta^{\gamma}_{\mu}$.\par
\indent
We would like to be able to write a transformation of the form

\begin{eqnarray}
\label{MINK1}
\Psi_{GKod}=e^{{i \over \hbar}\Gamma_{eff}[\vec\phi]}
\Bigl[e^{\theta\cdot\hat{T}}\Psi_{Kod}e^{-\theta\cdot\hat{T}}\bigr]
\end{eqnarray}

\noindent
were $\theta\sim\theta[\Gamma_{eff},\vec\phi]$ can be regarded as the parameter of the transformation, specified by the matter-specific model.  Note that the matter contribution to $\Psi_{GKod}$ can be written in the following `semiclassical' form

\begin{eqnarray}
\label{MINK2}
\Psi_{matter}[\vec{\phi}]=e^{{i \over \hbar}\Gamma_{eff}[\vec\phi]}=
e^{{i \over \hbar}\int_{\Sigma}d^3x\int_{\Gamma}\vec{f}\cdot{\delta}\vec{\phi}}
\end{eqnarray}

\noindent
In the absence of matter fields we would have $\Gamma_{eff}=\theta=0$.  So the question becomes, in what sense does the operator $\hat{T}$ act 
on $\Psi_{Kod}$.  We would like to interpret this operation as a kind of generalized matter-specific diffeomorphism of the CDJ 
matrix $\delta_{ae}$ corresponding to the pure Kodama state.  This matrix is invariant under $SO(3)$ rotations of the indices, in analogy to the Minkowski metric $\eta_{\mu\nu}$ begin invariant under Lorentz transformations.  However, we would like the action of a matter-induced transformation to produce a nontrivial effect, in analogy to (\ref{MINK}).  To isolate the CDJ matrix $\delta_{ae}$, it is convenient to invoke the instanton representation of the pure Kodama state derived in \cite{EYO}

\begin{eqnarray}
\label{MINK3}
\Psi_{Kod}[A]=e^{-6(\hbar{G}\Lambda)^{-1}\int_{\Sigma}({A}\wedge{dA}+(2/3){A}\wedge{A}\wedge{A})}
=e^{-6(\hbar{G}\Lambda)^{-1}\int_{M}\delta_{ab}{F^a}\wedge{F^b}}
\end{eqnarray}

\noindent
This leads to the following identity

\begin{eqnarray}
\label{MINK4}
(\Psi_{GKod})_{grav}=
\Bigl[e^{\theta\cdot\hat{T}}\Psi_{Kod}e^{-\theta\cdot\hat{T}}\bigr]
=e^{(\hbar{G})^{-1}\int_{M}\Psi_{ab}[\theta]{F^a}\wedge{F^b}}\nonumber\\
\hbox{exp}\Bigl[-6(\hbar{G}\Lambda)^{-1}\int_{M}
\Bigl(e^{\theta\cdot\hat{T}}\delta_{ab}e^{-\theta\cdot\hat{T}}\Bigr)
{F^a}\wedge{F^b}\Bigr].
\end{eqnarray}

\noindent
So we see that the effect of the operator $\hat{T}$ is to implement a transformation from the instanton representation of $\Psi_{Kod}$ into the gravitational sector of the instanton representation of $\Psi_{GKod}$.  The operation in the argument of the exponential in (\ref{MINK4}) can be written in terms of multiple commutators as a group transformation of the vector $\delta_{ab}$

\begin{eqnarray}
\label{COMMMU}
e^{\theta\cdot\hat{T}}\bigl[\delta_{ab}\bigr]e^{-\theta\cdot\hat{T}}
=\delta_{ab}+\theta^{a_1b_1}[\hat{T}_{a_1b_1},\delta_{ab}]
+{1 \over {2!}}\theta^{a_1b_1}\theta^{a_2b_2}
\bigl[\hat{T}_{a_1b_1},[\hat{T}_{a_2b_2},\delta_{ab}]\bigr]\nonumber\\
+{1 \over {3!}}\theta^{a_bb_1}\theta^{a_2b_2}\theta^{a_3b_3}
\bigl[\hat{T}_{a_1b_1}[\hat{T}_{a_2b_2},
[\hat{T}_{a_3b_3},\delta_{ab}]]\bigr]+\dots
\end{eqnarray}

\noindent
So in order to have the interpretation of $\Psi_{Kod}$ comprising an element of a vector space forming a representation of some group, it suffices to specify the realization of this group in this vector space.\footnote{This group action is unconvential in that the elements of the 
algebra $T_{ab}$ act on the `parameters' $\theta^{ab}$ as well as on the vector space of CDJ matrix elements.  In this respect in is a nonlinear group action.}\par
\indent
Recall from \cite{EYOFULL} that the CDJ matrix can be written in the following form

\begin{equation}
\label{MINK5}
\Psi_{ab}=-{6 \over \Lambda}\Bigl(\delta_{ab}+\sum_{n=1}^{\infty}(G\Lambda)^{n}
\hat{U}_{ab}^{a_{1}b_{1}a_{2}b_{2}...a_{n}b_{n}}\prod_{k=1}^{n}Q^{\prime}_{a_{k}b_{k}}\Bigr)=-{6 \over \Lambda}
\Bigl[e^{G\Lambda{Q}^{\prime}_{gh}T^{gh}}\Bigr]_{ab}^{ef}\delta_{ef}.
\end{equation}

\noindent
The idea is to express (\ref{MINK5}) as a group action implemented from the exponentiation of the appropriate generators by making the 
identification $\theta\equiv\theta_{ae}\sim{G}\Lambda{Q}^{\prime}_{ae}$.  We will see that the operator $T$ in (\ref{MINK5}) acts on $Q$ as well as on $\delta_{ef}$.  Let us expand the first few terms of the series.  

\begin{eqnarray}
\label{MINK6}
\Psi_{ab}=-{6 \over \Lambda}\Bigl[
\delta_a^e\delta_b^f
+(G\Lambda)\delta^{ef}T_{ab}^{a_1b_1}Q^{\prime}_{a_1b_1}\nonumber\\
+{{(G\Lambda)^2} \over {2!}}\delta^{ef}
T_{ae_2}^{a_1b_1}T_{e_2b}^{a_2b_2}Q^{\prime}_{a_1b_1}Q^{\prime}_{a_2b_2}
+{{(G\Lambda)^3} \over {3!}}\delta^{ef}
T_{ae_2}^{a_1b_1}T_{e_2e_3}^{e_2b_2}T_{e_3b}^{a_2b_2}
Q^{\prime}_{a_1b_1}Q^{\prime}_{a_2b_2}Q^{\prime}_{a_3b_3}\nonumber\\
+{{(G\Lambda)^4} \over {4!}}\delta^{ef}
T_{ae_2}^{a_1b_1}T_{e_2e_3}^{e_2b_2}T_{e_3e_4}^{a_2b_2}T_{e_4b}^{a_3b_3}
Q^{\prime}_{a_1b_1}Q^{\prime}_{a_2b_2}Q^{\prime}_{a_3b_3}Q^{\prime}_{a_4b_4}
\nonumber\\
+{{(G\Lambda)^5} \over {5!}}\delta^{ef}
T_{ae_2}^{a_1b_1}T_{e_2e_3}^{e_2b_2}T_{e_3e_4}^{a_2b_2}T_{e_4e_5}^{a_3b_3}
T_{e_5b}^{a_3b_3}
Q^{\prime}_{a_1b_1}Q^{\prime}_{a_2b_2}Q^{\prime}_{a_3b_3}Q^{\prime}_{a_4b_4}
Q^{\prime}_{a_5b_5}+\dots\Bigr]\delta_{ef}
\end{eqnarray}

\noindent
Let us now analyse the first few terms of (\ref{MINK6}).\par

\subsection{CDJ matrix as a cubic tree network series}

\noindent
The zeroth-order term of (\ref{MINK6}) merely reproduces the CDJ matrix for the pure Kodama state $\Psi_{Kod}$ as 
in $\delta_a^e\delta_b^f\delta_{ef}=\delta_{ab}$.  The first-order term represents the first-order (linearized) correction to $\Psi_{Kod}$ and involves the propagation of a disturbance from the source term $Q_{ab}$, which encodes the semiclassical limit through $\Gamma_{eff}$.  The operator 
$T_{ab}^{a_1b_1}$ is trivial in the sense that $Q^{\prime}_{ab}$ is the dimensionless propagated version of $Q_{ab}$.

\begin{eqnarray}
\label{MINK7}
T_{ab}^{a_1b_1}Q^{\prime}_{ab}
={1 \over 3}\delta_a^{a_1}\delta_b^{b_1}(O^{-1})_{a_1b_1}^{cd}Q_{cd}
={1 \over 3}\delta_a^{a_1}\delta_b^{b_1}R_{a_1b_1}^{a_1^{\prime}b_1^{\prime}}
(D^{-1})_{a_1^{\prime}b_1^{\prime}}^{cd}Q_{cd}.
\end{eqnarray}

\noindent
For the remaining terms we need only analyse the tensor product structure of the representation $T$.  The contribution to the second-order term of (\ref{MINK6}) is given by

\begin{eqnarray}
\label{MINK8}
T_{ae_2}^{a_1b_1}T_{e_2b}^{a_2b_2}
=\hat{U}_{ab}^{a_1b_1a_2b_2}
=U_{ab}^{gh}\Sigma_{gh}^{a_1b_1a_2b_2}
\end{eqnarray}

\noindent
The important thing to note concerning (\ref{MINK8}) is that $\Sigma$ corresponds to a trivalent node with the appropriate index structure to intertwine the representation of the propagating mode of the disturbance from matter source $Q_{ab}$ with the representation of the vertex such such that there is conservation of a `generalized' angular momentum associated with the generator $T$.  The index structure is set up to absorb the disturbance propagated from $Q_{a1b1}$ and $Q_{a_2b_2}$, and retransmit it through the indices $gh$ to the point $x$.\par
\indent
The third-order term of (\ref{MINK6}) is the tetravalent node and consists of three terms, given by

\begin{eqnarray}
\label{MINK9}
T_{ae_2}^{a_1b_1}T_{e_2e_3}^{e_2b_2}T_{e_3b}^{a_2b_2}
=\hat{U}_{ab}^{a_1b_1a_2b_2a_3b_3}
=U_{ab}^{gh}E_{gh}^{a_1b_1a_2b_2a_3b_3}\nonumber\\
+U_{ab}^{gh}\Sigma_{gh}^{g_1h_1g_2h_2}
U_{g_1h_1}^{g_3h_3}\Sigma_{g_3h_3}^{a_1b_1a_2b_2}U_{g_2h_2}^{a_3b_3}
+U_{ab}^{gh}\Sigma_{gh}^{g_1h_1g_2h_2}U_{g_1h_1}^{a_1b_1}
U_{g_2h_2}^{ef}\Sigma_{ef}^{a_2b_2a_3b_3}.
\end{eqnarray}

\noindent
The first term is designed to absorb the disturbances transmitted from three sources $Q_{a_1b_1}$, $Q_{a_2b_2}$ and $Q_{a_3b_3}$, rotating them into the correct representation to link with the E node, which then retransmits them through the $gh$ index.  The second and third terms are the two permutations for a pair of charges to propagate a disturbance which combines at a trivalent node, which propagates in turn to combine with the disturbance from a single source, the resultant of which propagates onward to $x$.\par
\indent
The fourth-order term begins to illustrate the workings of the network.  This can be written in the form

\begin{eqnarray}
\label{MINK10}
T_{ae_2}^{a_1b_1}T_{e_2e_3}^{a_2b_2}T_{e_3e_4}^{a_3b_3}T_{e_4b}^{a_4b_4}
=\hat{U}_{ae_3}^{a_1b_1a_2b_2}\hat{U}_{e_3b}^{a_3b_3a_4b_4}
+\hat{U}_{ae_2}^{a_1b_1}\hat{U}_{e_2b}^{a_2b_2a_3b_3a_4b_4}
+\hat{U}_{ae_4}^{a_1b_1a_2b_2a_3b_3}\hat{U}_{e_4b}^{a_4b_4}\nonumber\\
=U_{ab}^{gh}U_{ge_3}^{g_1h_1}U_{e_3h}^{g_2h_2}
\Sigma_{h_1g_1}^{a_1b_1a_2b_2}\Sigma_{h_2g_2}^{a_3b_3a_4b_4}
+U_{ab}^{gh}\Sigma^{e_1f_1e_2f_2}_{gh}U_{e_2f_2}^{a_4b_4}
U_{e_1f_1}^{g_1h_1}E_{g_1h_1}^{a_1b_1a_2b_2a_3b_3}\nonumber\\
+U_{ab}^{gh}\Sigma_{gh}^{g_1h_1g_2h_2}U_{g_2h_2}^{a_1b_1}
E_{g_1h_1}^{a_2b_2a_3b_3a_4b_4}
\end{eqnarray}

\noindent
In the first term of (\ref{MINK10}) a flowpath is set up to absorb disturbances from two pairs of sources, each pair propagating into a trivalent vertex.  The two trivalent vertices then propagate the disturbance on to recombine at another trivalent vertex, which in turn propagates it to the point $x$.  This can only occur, due to the index structure, in one way.  The second and third terms are permutations on the manner in which a triplet of sources propagates a disturbance into a trivalent vertex which transmits the disturbance ot combine with a disturbance from a single source, the combination occuring at a trivalent vertex which in turn propagates to the point $x$\par
\indent
A typical contribution to the fifth-order term of (\ref{MINK6}) is given by

\begin{eqnarray}
\label{MINK11}
T_{ae_2}^{a_1b_1}T_{e_2e_3}^{a_2b_2}T_{e_3e_4}^{a_3b_3}T_{e_4e_5}^{a_4b_4}
T_{e_5b}^{a_5b_5}
=\hat{U}_{aa_3}^{a_1b_1a_2b_2}\hat{U}_{a_3b}^{a_3b_3a_4b_4a_5b_5}
+\hat{U}_{aa_4}^{a_1b_1a_2b_2a_3b_3}\hat{U}_{a_4b}^{a_4b_4a_5b_5}\nonumber\\
=U_{ab}^{gh}U_{ga_3}^{g_1h_1}U_{a_3h}^{g_2h_2}
\Sigma_{g_1h_1}^{a_1b_1a_2b_2}E_{g_2h_2}^{a_3b_3a_4b_4a_5b_5}
+U_{ab}^{gh}U_{ga_4}^{g_1h_1}U_{a_4h}^{g_2h_2}
E_{g_1h_1}^{a_1b_1a_2b_2a_3b_3}\Sigma_{g_2h_2}^{a_4b_4a_5b_5}.
\end{eqnarray}

\noindent
Equation (\ref{MINK11}) is designed to receive disturbances from five sources $Q_{ab}$.  According to the index structure, these disturbances distribute themselves into a pair and a triplet, which rotate into a trivalent and tetravelent vertex respectively, the outputs of which combine into a trivalent node which propagates the disturbance on to the point $x$.  This can occur in two ways, which is a constraint imposed by network topology.  Let us move on to the sixth-order term, for illustrative purposes.  A typical term is given by

\begin{eqnarray}
\label{MINK12}
T_{ae_2}^{a_1b_1}T_{e_2e_3}^{a_2b_2}T_{e_3e_4}^{a_3b_3}T_{e_4e_5}^{a_4b_4}
T_{e_5e_6}^{a_5b_5}T_{e_6b}^{a_6b_6}\nonumber\\
=\hat{U}_{ae_3}^{a_1b_1a_2b_2}\hat{U}_{e_3e_5}^{a_3b_3a_4b_4}
\hat{U}_{e_5b}^{a_5b_5a_6b_6}
+\hat{U}_{ae_4}^{a_1b_1a_2b_2a_3b_3}\hat{U}_{e_4b}^{a_4b_4a_5b_5a_6b_6}
\nonumber\\
=U_{ab}^{hg}E_{hg}^{h_1g_1h_2g_2h_3g_3}
U_{h_1g_1}^{e_1f_1}\Sigma_{e_1f_1}^{a_1b_1a_2b_2}.
\end{eqnarray}

\noindent
The $n^{th}$ order term is a sum over the number of different network topologies connecting a set of $n$ sources through (model-specific) trivalent and tetravalent nodes to the point $x$, given in general by

\begin{eqnarray}
\label{MINK13}
(T\dots{T})^{a_1b_1\dots{a}_nb_n}_{ab}
=\sum_{k,l}\Bigl(\prod_{k=0}^K\Sigma_{e_kf_k}^{a_kb_kc_kd_k}
\prod_{l=K}^{n}E_{g_lh_l}^{a_lb_lc_ld_le_lf_l}
\prod_n{U}_{m_dn_d}^{p_dq_d}\Bigr)
\end{eqnarray}

\noindent
The main purpose of this section is to motivate a kind of correspondence between the network description of gravity, which can be seen as the perturbative solution to the classical equations of motion arising from the quantum constraints a second-quantized theory, and the quantum treatment of a third-quantized theory, the theory of quantum fluctuations on DeSitter spacetime.  There is a direct correspondence since the latter quantum theory (\ref{MATTTEER}) and the classical 
counterpart (\ref{MINK13}) contain the same structures, namely the vertices (or nodes) $V_2$ and $V_3$, as well as the propagators and source (matter) currents.  A similar analogy can be drawn for 
scalar $\phi^4$ and Yang--Mills theories, two renormalizable theories.  This suggests the inference of gravity as a renormalizable quantum theory in Ashtekar variables if one makes the 
identification $G\Lambda\sim{g}$, where $g$ is the dimensionless Yang--Mills coupling constant.  Note that $G\Lambda$ is as well dimensionless.

\section{Discussion}

\noindent
In a usual quantum theory, one expands the theory about a vacuum configuration of the fields and then attempts to make predictions which might be testable by experiment.  In some cases when the wrong vacuum is selected, the quantum theory turns out to be nonrenormalizable.  This is definitely the case in metric general relativity when expanding about Minkowski spacetime, owing to the negative mass dimension of the associated coupling constant $G$.  However, in the Ashtekar variables, it appears that the DeSitter spacetime with associated cosmological term $\Lambda$ allows for the possiblity of a renormalizable expansion.  This is due to the zero mass dimension of its associated coupling 
constant $G\Lambda$ when viewed as a quantum theory, as well as the polynomial form of the theory which the metric variables lack.  Still, it is a general result of quantum theory that even if a theory may be renormalizable, it may be the case that the perturbative treatment, when carried out to all orders, may still be divergent thus necessitating considerations of Borel summability of the perturbative series.  Our interpretation of the generalized Kodama states $\Psi_{GKod}$ is that it is the direct analogue of when one expands a perturbatively renormalizable theory to all orders in the expansion.  The finiteness of this expansion is an area to be examined in greater depth.  As a good example, $\phi^4$ theory is renormalizable and is considered finite, yet its effective action forms a divergent hypergeometric series expansion in the coupling parameters of the theory.  We will ultimately examine in future works whether or not this is the case for the generalized Kodama states $\Psi_{GKod}$ in relation to $\Psi_{Kod}$.  It is clear the two wavefunctions are nonperturbatively related to all orders in an expansion of the former about the latter.  In this paper we have provided a specific form for the generator of a kind of discrete transformation relating the two states in terms of tree networks.  The next step along this line of reasoning will be to demonstrate in greater detail how one can cast this into in terms of a renormalizable theory of quantum gravity.  The basis of the argument is that there exist two renormalizable theories in four 
dimensions, scalar $\phi^4$ and Yang--Mills theory, whose classical equations of motion have the same structure as the constraints $C_{ae}=0$ of relativity in Ashtekar variables.  Therefore these three theories should possess networks of identical structure and topology.  From this perspective, when one relates the network structure to the renormalizability properties of the two former theories, then one should be able to infer the renormalizability of gravity by isomorphism.  We will demonstrate this explicitly in a separate paper, as one of the future directions of research.


\begin{thebibliography}{99}

\bibitem{EYO} {Eyo Ita `Finite states in four dimensional quantized gravity',
arXiV:gr-qc/0703052 (To appear in Class. Quant. Grav. journal)}

\bibitem{EYOFULL} {Eyo Ita `A systematic approach to the solution of the constraints of quantum gravity: The full theory.' 
arXiV:gr-qc/0710.2364}

\bibitem{SOO} {Chopin Soo and Lee Smolin `The Chern--Simons invariant as the natural time variable for classical and quantum cosmology' 
Nucl.Phys.B449(1995)289-316}

\bibitem{LINKO} {Laurent Friedel and Lee Smolin 'The linearization of the Kodama state'
Class.Quantum Grav. 21(2004)3831-3844}

\bibitem{ASH} {Ahbay Ashtekar, Joseph D. Romano, and Ranjeet S. Tate `New variables for
gravity: Inclusion of matter' Pys. Rev. D40 (1989)2572}

\bibitem{ASH1} {Ahbay Ashtekar. `New perspectives in canonical gravity', (Bibliopolis, Napoli, 1988).}

\bibitem{ASH2} {Ahbay Ashtekar `New Hamiltonian formulation of general relativity'
Phys. Rev. D36(1987)1587}

\bibitem{ASH3} {Ahbay Ashtekar `New variables for classical and quantum gravity'
Phys. Rev. Lett. Volume 57, number 18 (1986)}

\bibitem{EYOPATH} {Eyo Ita `Finite states in 4 dimensional quantized gravity. A brief introduction into the path integration approach in Ashtekar variables' 
arXiV:0804.0793[gr-qc]}

\bibitem{EYOGAUSS} {Finite states of four dimensional quantized gravity. General solution to the Gauss' law constraint. 
arXiV:0704.0367[gr-qc]}

\bibitem{DEGENERATE} {Gary T Horowitz `Topology change in classical and quantum gravity', Class. Quantum Grav. 8(1991) 587-601}

\bibitem{EYOCONS} {Eyo Ita 'Generalized Kodama states as a unified description of gravity quantum gravity and matter fields'
In preparation}





\end{thebibliography}
\end{document}